\documentclass[11pt]{article}

\usepackage{amsfonts,amsmath,graphicx}

\usepackage{algpseudocode}
\usepackage{algorithm}
\newcommand{\pushcode}[1][1]{\hskip#1\dimexpr\algorithmicindent\relax}
\floatname{algorithm}{Transducer}

\usepackage{verbatim}

\usepackage[margin=1in]{geometry}

\begin{document}
\newcommand{\dom}{\mbox{$\rightarrow$}}
\newtheorem{tm}{\hspace{6mm}Theorem}[section]
\newtheorem{prp}[tm]{\hspace{6mm}Proposition}
\newtheorem{dfn}[tm]{\hspace{6mm}Definition}
\newtheorem{obs}[tm]{\hspace{6mm}Observation}
\newtheorem{claim}[tm]{\hspace{6mm}Claim}
\newtheorem{lemma}[tm]{\hspace{6mm}Lemma}
\newtheorem{cor}[tm]{\hspace{6mm}Corollary}
\newtheorem{conj}[tm]{\hspace{6mm}Conjecture}
\newtheorem{prob}[tm]{\hspace{6mm}Problem}
\newtheorem{quest}[tm]{\hspace{6mm}Question}
\newtheorem{alg}[tm]{\hspace{6mm}Algorithm}
\newtheorem{sub}[tm]{\hspace{6mm}Algorithm}
\newcommand{\ndom}{\mbox{$\not\rightarrow$}}
\newcommand{\lr}{\mbox{$\longrightarrow$}}

\newcommand{\qed}{\hfill$\diamond$}
\newcommand{\pf}{\vspace{-3mm}\hspace{6mm}{\bf Proof: }}
\newcommand{\nl}{\hfill\break}
\newcommand{\emset}{\emptyset}
\newcommand{\lefa}{\leftarrow}
\newcommand{\bbB}{\mathbb{B}}
\newcommand{\bbA}{\mathbb{A}}

\title{Space complexity of list $H$-colouring: a dichotomy}

\date{}

\author{
L\'{a}szl\'{o} Egri\thanks{Institute for Computer Science and Control, Hungarian Academy of Sciences (MTA SZTAKI), Budapest, Hungary, \texttt{laszlo.egri@mail.mcgill.ca}}
\and
Pavol Hell\thanks{School of Computer Science, Simon Fraser University, Burnaby, Canada, \texttt{pavol@sfu.ca}}
\and
Benoit Larose\thanks{Department of Mathematics and Statistics, Concordia University, Montreal, Canada, \texttt{benoit.larose@concordia.ca}}
\and
Arash Rafiey\thanks{School of Computer Science, Simon Fraser University, Burnaby, Canada, \texttt{arashr@sfu.ca}}
}

\maketitle

\begin{abstract}
The Dichotomy Conjecture for constraint satisfaction problems (CSPs) states that every CSP is in P or is NP-complete (Feder-Vardi, 1993). It has been verified for conservative problems (also known as list homomorphism problems) by A. Bulatov (2003). We augment this result by showing that for digraph templates $H$, every conservative CSP, denoted LHOM($H$), is solvable in logspace or is hard for NL. More precisely, we introduce a digraph structure we call a circular $N$, and prove the following dichotomy: if $H$ contains no circular $N$ then LHOM($H$) admits a logspace algorithm, and otherwise LHOM($H$) is hard for NL. Our algorithm operates by reducing the lists in a complex manner based on a novel decomposition of an auxiliary digraph, combined with repeated applications of Reingold's algorithm for undirected reachability (2005). We also prove an algebraic version of this dichotomy: the digraphs without a circular $N$ are precisely those that admit a finite chain of polymorphisms satisfying the Hagemann-Mitschke identities. This confirms a conjecture of Larose and Tesson (2007) for LHOM($H$). Moreover, we show that the presence of a circular $N$ can be decided in time polynomial in the size of $H$. 
\end{abstract}

\section{Introduction}

Fixed-template constraint satisfaction problems (CSPs) provide a unifying framework for a wide range of natural problems
arising both in applied and theoretical computer science. Examples include 3-SAT, HORN-3-SAT, 2-colouring, directed and undirected reachability, linear equations on a finite field, etc. Since T.\ Feder and M.\ Vardi stated their celebrated dichotomy conjecture in their seminal 1993 paper \cite{fedvar}, predicting that every CSP is either polynomial-time tractable or NP-complete \cite{fedvar}, CSPs have been the subject of intense scrutiny from the perspective of computational complexity. Indeed, fuelled mostly by the injection of tools borrowed from universal algebra, the last two decades have witnessed a major progress in the field.

Among the fruits of the marriage of CSP-theory and universal algebra, there are two deep conjectures \cite{lartes, larzad}, proposing a more refined computational and descriptive complexity classification of CSPs. In fact, the first one, the so-called bounded width conjecture, has been settled recently by Barto and Kozik \cite{barkoz}. Essentially, this beautiful result describes, in algebraic terms, those CSPs that can be solved by local-consistency methods, or equivalently, those expressible in the logic programming language Datalog. The second conjecture, due to Larose and Tesson \cite{lartes}, postulates a dichotomy within this class of CSPs: every CSP expressible in Datalog is either solvable in logspace or is hard for NL.

Already in their 1993 paper, Feder and Vardi showed that proving the dichotomy conjecture for CSPs with digraph templates is sufficient to prove it in its full generality. In fact, it is shown in \cite{jdmt} that for every CSP($\bbB$), where $\bbB$ is a general template, there is a digraph template $H_{\bbB}$ such that CSP($\bbB$) and CSP($H_{\bbB}$) are logspace equivalent. That is, to study the refined computational complexity of CSPs, it is sufficient to focus on digraph templates.

One of the early successes of the algebraic method is the proof of the Feder-Vardi dichotomy conjecture for general structures in the \emph{conservative} case \cite{bul,bul_journal}. This result was motivated by a specific dichotomy classification for the case of undirected graphs \cite{biarc}. In \cite{soda}, the specific classification was extended to the case of digraphs, thereby refining the dichotomy from \cite{bul,bul_journal} by specifying which conservative CSPs with digraph templates $H$ (or equivalently which list homomorphism problems LHOM($H$)) are tractable or intractable by means of a forbidden substructure called a {\em digraph asteroidal triple (DAT)} (see Definition \ref{def-dat}). More precisely, the authors show that LHOM($H$) is in P if $H$ contains no DAT, and is NP-complete otherwise. In fact, Hell and Rafiey's work also shows that either LHOM($H$) has bounded width (see also \cite{kaz}), or it is NP-complete. Recalling the result of Barto and Kozik (the technicalities also rely on \cite{hobmc}), now we are ready to state the refined conjecture of Larose and Tesson in the context of list-homomorphism problems with digraph templates: either LHOM($H$) can be solved in logspace, or it is hard for NL. The main result of this paper is a proof of this conjecture.

\textbf{Our results:} For ease of presentation, we have chosen to keep the algebra to a minimum and focus our attention on the combinatorial aspects of our results. In addition, we restrict our definitions in this paper to CSPs with digraph templates, but we note that these definitions can be easily extended to CSPs with general templates.

An instance of LHOM($H$) is a pair $(G,L)$, where $G$ is a digraph and $L$ is a collection of lists $L(x) \subseteq V(H), x \in V(G)$. A {\em list-homomorphism} $f$ of $G$ to $H$ with respect to $L$, also called an $L$-homomorphism of $G$ to $H$, is a mapping $f : V(G) \to V(H)$ which is a homomorphism, i.e., has $f(x)f(y) \in E(H)$ whenever $xy \in E(G)$, and which respects the lists, i.e., has $f(x) \in L(x), x \in V(G)$. The task is to decide whether or not $G$ admits an $L$-homomorphism to $H$.\footnote{ LHOM$(H)$ is identical to CSP($\bbB$) where $\bbB$ is a relational structure that contains the set of arcs of the digraph and all subsets of $H$.}

An arc-preserving map $f:H^n \rightarrow H$ is called a {\em polymorphism} of the digraph $H$. The algebraic approach mentioned above relies on the simple but deep observation that the computational complexity of CSP($H$) is controlled by the properties of the polymorphisms of $H$ \cite{BJK,bulval,JCG,lartes}. In their proof of the bounded width conjecture, Barto and Kozik characterise via a precise set of \emph{identities} (polymorphisms satisfying certain equations) those CSPs solvable by local-consistency methods. As mentioned before, it is believed \cite{hobmc,lartes} that within this family of CSPs, those whose template admits a so-called finite \emph{chain of Hagemann-Mitschke (HM-chain) polymorphisms} (see Definition~\ref{def-HM}) are solvable in logspace, and all other CSPs are hard for NL.

We characterise digraphs admitting an HM-chain of polymorphisms via a forbidden structure we call a {\em circular $N$} (Definition~\ref{def-cN}). Moreover, we prove that they are precisely those digraphs for which the list homomorphism problem is in logspace. We also show that the presence of a circular $N$ in a digraph $H$ can be detected in polynomial time. The main results of the paper can be summarized as follows.

\begin{tm}\label{main_thm}
Let $H$ be a digraph. Then $H$ admits a finite HM-chain of polymorphisms if and only if $H$ has no circular $N$.
If one of these equivalent conditions is satisfied, then there is a logspace algorithm that decides LHOM($H$), and otherwise
LHOM($H$) is NL-hard. Furthermore, the presence of a circular $N$ in $H$ can be tested in time polynomial in $|V(H)|$.
\end{tm}

The most challenging part of the above theorem is devising an algorithm that solves LHOM($H$) in logspace when $H$ contains no circular $N$. In fact, one of the main difficulties faced today by researchers studying the complexity of CSPs is the paucity of available efficient algorithms; we believe that the algorithm we describe here is a non-trivial contribution to this research area.

Prior to our results, the only known complexity-theoretic consequences of the existence of an HM-chain of polymorphisms were, for general CSPs, a proof by Bulatov and Dalmau \cite{buldal} that an HM-chain of polymorphisms of length {\em one} implies polynomial-time tractability;  and Dalmau and Larose \cite{dallar} showed that combined with Datalog, a chain of length one implies solvability in logspace. In the present paper we prove that for LHOM($H$), an HM-chain of polymorphisms of {\em any} length implies solvability in logspace.

In general, it is known that structures that do not admit an HM-chain of polymorphisms have an associated CSP which is NL-hard \cite{lartes}. However, in the context of LHOM($H$), our results show that the only reason LHOM($H$) is NL-hard is that the 2-element order relation can be defined from $H$ via a very simple kind of \emph{primitive-positive definition}. In turn, this yields a simple logspace reduction from directed unreachability to LHOM($H$).

Finally, we remark that our graph-theoretic results might be of independent interest to graph theorists, as they generalize previous results for undirected graphs (see \cite{eklt,yachecha}). In particular, the class of digraphs that contain no circular $N$ generalizes the class of trivially perfect undirected graphs, where each vertex has a loop, as well as the class of $(P_6,C_6)$-free bipartite graphs.

\textbf{Our techniques and outline:} In Section \ref{sect-prelim}, we introduce basic terminology and notation, and define the notion of circular $N$. Theorem \ref{circN-nl-hard} confirms that if $H$ has a circular $N$ then LHOM($H$) is NL-hard. The section introduces the key notion of the auxiliary pair digraph $H^+$, and prove various basic facts about it.

Section \ref{sect-algo} contains the description of the logspace algorithm and in Section 4, we prove the correctness of the algorithm. We note that in the special case of graphs, an inductive graph-theoretic characterisation of the graphs $H$ that admit an HM-chain of polymorphisms is known \cite{eklt}, and this characterisation can be used to inductively construct a simple algorithm for LHOM($H$). However, when $H$ is a digraph, no such characterisation is available, and consequently, our algorithm for problems with digraph templates becomes much more involved.

The basic idea of the algorithm is as follows. Let LHOM$_s$($H$) denote the problem LHOM($H$) where each list has size at most $s$. The algorithm is constructed inductively: assuming that we have a logspace algorithm that solves LHOM$_{s-1}$($H$), we construct a logspace algorithm for LHOM$_s$($H$). Since LHOM$_1$($H$) can be solved easily and LHOM$_h$($H$), where $h = |V(H)|$, is just LHOM($H$), once the induction step is established, we obtain the desired algorithm. The heart of this induction step is a logspace transducer $T(a,b)$. The transducer $T(a,b)$ takes as input a digraph $G$ with a collection of lists $L$ (a list for each vertex of $G$), and for each $v \in V(G)$ such that $a,b \in L(v)$, removes either $a$ or $b$ from $L(v)$, such that the new instance has a list homomorphism to $H$ if and only if the original instance had a list-homomorphism to $H$. The transducer relies on LHOM$_{s-1}$($H$), and it heavily employs Reingold's algorithm for undirected connectivity \cite{reingold} to test different reachability conditions in the underlying graph of a certain auxiliary triple digraph. The correctness of $T(a,b)$ crucially relies on the absence of circular $N$s in $H$. Finally, the algorithm for LHOM$_s$($H$) is a sequence of the transducers $T(a_m,b_m),T(a_{m-1},b_{m-1}),\dots,T(a_1,b_1)$ chained together, where $(a_j,b_j)$ are defined based on the structure of $H^+$.

We note that the main challenge when constructing $T(a,b)$ is to do all the computations in logspace. Indeed, if linear space is available, then all lists can be written down simultaneously on the tape of the Turing machine, and this allows us to construct a significantly simpler polynomial time algorithm.

Section \ref{sect-nice-terms} is devoted to proving the logspace conjecture of \cite{lartes} per se: we define the notion of a Hagemann-Mitschke chain, and show in Theorem \ref{thm51} that a digraph admits such a chain of polymorphisms if and only if it has no circular $N$.
In fact, we prove that we can determine in polynomial time the least $k$, if any, for which a digraph admits an HM-chain of polymorphisms of length $k$; in particular we can check if a digraph has a circular $N$ efficiently, which we prove in section \ref{sect-additional} (Theorem \ref{circN-recog}). This last section also contains additional remarks about FO-definable LHOM($H$) and some open problems.
\section{Preliminaries} \label{sect-prelim}

A {\em digraph} $H$ consists of a set $V(H)$ of {\em vertices} and a set $E(H)$ of {\em arcs}. An arc $uv \in E(H)$ will also be called a {\em forward edge} of $H$; moreover, $uv$ will be called a {\em backward edge} of $H$ if $vu \in E(H)$. (Thus a forward edge is just an arc, and a backward edge is a reversed arc.) An edge is {\em single} if it is either forward or backward, and {\em double} if it is both forward and backward; similarly, an arc is single if it is a single forward edge. A walk is {\em directed} (forward or backward) if all its edges are in the same direction (forward or backward respectively); note that these edges can be single or double. If $X = x_0, x_1, \dots, x_n$ is a walk, we define the {\em reversed} walk to be $X^{-1} = x_n, \dots, x_1$. Note that forward edges of $X$ become backward edges of $X^{-1}$ and vice versa.

We define two walks $X = x_0, x_1, \dots, x_n$ and $Y = y_0, y_1, \dots, y_n$ in $H$ to be {\em congruent}, if they follow the same pattern of forward and backward edges, i.e., if $x_ix_{i+1}$ is a forward edge if and only if $y_iy_{i+1}$ is a forward edge. Suppose $X, Y$ and $Z = z_0,$ $z_1, \dots, z_n$ are congruent walks. We say that $x_iy_{i+1}$ is a {\em faithful edge from $X$ to $Y$} if it is an edge of $H$ in the same direction (forward or backward) as $x_ix_{i+1}$. We say that $X$ {\em avoids} $Y$ in $H$ if there is no faithful edge from $X$ to $Y$ in $H$. Note that $X$ avoids $Y$ if and only if $Y^{-1}$ avoids $X^{-1}$. Observe that two walks of length zero also avoid each other.

We say that $Z$ {\em protects} $Y$ from $X$ if the existence of faithful edges $x_iz_{i+1}$ and $z_jy_{j+1}$ in $H$ implies that $j \leq i$. In other words, $Z$ protects $Y$ from $X$ if and only if there exists a subscript $s$ such that $x_0, x_1, \dots, x_s$ avoids $z_0, z_1, \dots, z_s$ and $z_{s+1}, z_{s+1}, \dots, z_n$ avoids $y_{s+1}, y_{s+2}, \dots, y_n$.

An \emph{invertible pair} in $H$ is a pair of vertices $u,v$, such that
\begin{itemize}
\item there exist congruent walks $P$ from $u$ to $v$ and $Q$ from $v$ to $u$, such that $P$ avoids $Q$,
\item and there exist congruent walks $P'$ from $v$ to $u$ and $Q'$ from $u$ to $v$, such that $P'$ avoids $Q'$.
\end{itemize}

A digraph with vertices $x, x', y, y'$ and edges $xx', yy', yx'$ (all in the same direction, forward or backward) is called an $N$.

\begin{dfn}\label{def-cN}
Let $x, x', y, y'$ be vertices of a digraph $H$. An {\em extended $N$ from $x,x'$ to $y,y'$} in $H$ consists of congruent walks $X$ (from $x$ to $x'$), $Y$ (from $y$ to $y'$), and $Z$ (from $y$ to $x'$), such that $X$ avoids $Y$ and $Z$ protects $Y$ from $X$. A {\em circular $N$} is an extended $N$ in which $x = x'$ and $y = y'$.
\end{dfn}

The following result follows directly from Lemma \ref{cN-noHM} and algebraic results in \cite{lartes}. For the sake of completeness, we include a short direct proof in the case of digraph list homomorphism problems.

\begin{tm} \label{circN-nl-hard}
If $H$ contains a circular $N$, then LHOM$(H)$ is NL-hard.
\end{tm}

\pf
We give a logspace reduction of directed $st$-connectivity to LHOM$(H)$. Suppose there is a circular $N$, with congruent walks $X$, consisting of $x_0=x, x_1, \dots, x_n=x$, $Y$ consisting of $y_0=y, y_1, \dots, y_n=y$, and $Z$, consisting of $z_0=y, z_1, \dots, z_n=x$. Define a path $P$ with vertices $p_0, p_1, \dots, p_n$ and (forward or backward) edges $p_ip_{i+1}$ so that $P$ is congruent to $X$ (and hence also to $Y$ and $Z$). Define the lists $L(p_i) = \{x_i, y_i, z_i\}$. It now follows that there is a list homomorphism $f$ of $P$ to $H$ with $f(p_0)=x, f(p_n)=x$, a list homomorphism $g$ of $P$ to $H$ with $g(p_0)=y, f(p_n)=y$, and a list homomorphism $h$ of $P$ to $H$ with $h(p_0)=y, h(p_n)=x$. It is also easy to see that there is no list homomorphism of $P$ to $H$ taking $p_0$ to $x$ and $p_n$ to $y$, because $X$ avoids $Y$ and $Z$ protects $Y$ from $X$.

Given a graph with vertices $s$ and $t$, we construct a graph $G'$ by replacing each arc $uv$ of $G$ by a copy of the path $P$, denoted by $p_0', p_1', \dots, p_n'$, and identifying vertex $u$ with $p_0'$, and vertex $v$ with $p_n'$. The list of each $p_i'$ is $L(p_i') = \{x_i, y_i, z_i\}$. Furthermore, we add the lists $L(s) = \{x\}$, and $L(t) = \{y\}$. It is easy to check that there is an $L$-homomorphism from $G'$ to $H$ if and only if there is no path from $s$ to $t$ in $G$. It is easy to check that the reduction can be carried out in logspace.
\qed

\begin{lemma}\label{helpful}
If $H$ contains an extended $N$ with walks $X, Y, Z$ such that $Y$ also avoids $X$, then $H$ contains a circular $N$.
\end{lemma}

\pf
If there are no faithful edges from $Y$ to $X$ then we obtain a circular $N$ by taking $X, Y, Z$ and following them by the reversed walks $X^{-1}, Y^{-1}$ and $X^{-1}$ respectively.
\qed

Let $H$ be a digraph. We define the following {\em pair digraph} $H^+$. The vertices of $H^+$ are all ordered pairs $(x,y)$, where $x, y$ are distinct vertices of $H$. There is an arc from $(x,y)$ to $(x',y')$ in $H^+$ in one of the following situations:

\begin{itemize}
\item
$xx' \in E(H), yy' \in E(H), xy' \not\in E(H)$, or
\item
$x'x \in E(H), y'y \in E(H), y'x \not\in E(H)$.
\end{itemize}

Note then an $N$ with vertices $x, x', y, y'$ in $H$ corresponds precisely to a single arc of $H^+$. A double arc of $H^+$ corresponds to vertices $x, x', y, y'$ and edges $xx', yy'$ in $H$ such that $xy', yx'$ are not edges in $H$ (all in the same direction, forward or backward).

\begin{lemma}\label{N}
Assume $H$ contains no circular $N$. If $C$ is a directed cycle in $H^+$, then all arcs of $C$ are double.
\end{lemma}

\pf
Suppose $C = (x_0,y_0), (x_1,y_1), \dots, (x_t,y_t),(x_0,y_0)$ is a directed cycle in $H^+$. Assume without loss of generality that $(x_0,y_0)(x_1,y_1)$ is a single (forward or backward) arc in $H^+$. Let $X$ be the walk $x_0,x_1,\dots,x_t,x_0$ and $Y$ be the walk $y_0,y_1,\dots,y_t,y_0$. Observe that $X$ avoids $Y$. Let $Z$ be the walk $y_0,x_1,x_2,\dots,x_t,x_0$. It is easy to see that $Z$ protects $Y$ from $X$. Therefore $X$, $Y$, and $Z$ form a circular $N$.
\qed

\begin{cor}\label{strong_double}
All arcs in a strong component of $H^+$ are double.
\end{cor}

We shall {\bf assume from now on that $H$ is a fixed digraph that does not contain a circular $N$}.

The {\em condensation} $C(H^+)$ of $H^+$ is obtained from $H^+$ by replacing each strong component of $H^+$ by a single vertex, with an arc from $u$ to $v$ ($u$ different from $v$) if there is any arc from any vertex $(x,y)$ of the strong component  corresponding to $u$ to any vertex $(x',y')$ of the strong component corresponding to $v$. A strong component of $H^+$ corresponding to a vertex of $C(H^+)$ with no out-going (in-coming) arcs  is called a {\em sink component} (respectively a {\em source component}). For each strong component $C$ there is a {\em reverse strong component} $C'=\{(y,x) | (x,y) \in C\}$; it is easy to check that the reverse component is in fact a strong component of $H^+$. (The reason for this is the {\em skew property} of $H^+$ that says that if $(x,y)(x',y')$ is an arc of $H^+$, then $(y',x')(y,x)$ is also an arc of $H^+$.) It is possible that $C'=C$, in which case each pair $(x,y) \in C$ is {\em invertible}, i.e., has also $(y,x) \in C$. Moreover, if $C$ is a sink component, then $C'$ is a source component and vice versa. Since the condensation of any digraph is acyclic, it contains a source and sink. Therefore, we can order the strong components of $H^+$ as $C_1, C_2, \dots, C_t$, so that $C_i$ is a sink component of $H^+ - C_{i+1} - C_{i+2} - \dots - C_t$, i.e., has no arcs to $C_j, j < i$. We fix an ordering of the vertices of $H^+$, as $p_1, p_2, \dots, p_m$, in which we consecutively list the vertices of $C_1$, then $C_2$, etc., until the vertices of $C_t$, arbitrarily ordered inside each $C_i$. We remark that in this ordering, if $p_i$ has a single arc to $p_j$ then $p_i$ and $p_j$ are in different strong components $C_{i'}$ and $C_{j'}$ respectively, with $j' > i'$. (In particular, also $j > i$.) Note for future reference that the component which contains a pair $p_k$ is denoted by $C_{k'}$. Recalling that the vertices $p_i$ of $H^+$ are ordered pairs of distinct vertices of $H$, let us suppose each $p_i=(a_i,b_i)$. Then we obtain the following equivalent condition we will need later.

\begin{itemize}
\item[$(*)$]
If $a_ia_j$, $b_ib_j$, and $b_ia_j$ are forward (respectively backward) edges of $H$ while $a_ib_j$ is not a forward (respectively backward) edge of $H$, and if $(a_i,b_i)$ is in the strong component $C_{i'}$, and $(a_j,b_j)$ in the strong component $C_{j'}$, then we must have $j' > i'$.
\end{itemize}

For a fixed $k$, we say that the pair $p_i$ is \emph{$k$-processed}, if $i > k$. An unordered pair $u,v$ is \emph{$k$-allowed} if neither $(u,v)$ nor $(v,u)$ are $k$-processed. We say that the collection $L$ of lists is {\em $k$-good} if for every $x \in V(G)$ and any distinct $u, v \in L(x)$, the unordered pair $u,v$ is $k$-allowed. 
\section{The Logspace Algorithm} \label{sect-algo}

The goal of this section is to provide a logspace algorithm whose existence is claimed in Theorem~\ref{main_thm}. We denote by LHOM$_s$($H$) the restriction of LHOM($H$) to inputs $(G,L)$, such that for each $v \in V(G)$, $|L(v)| \leq s$. We inductively construct a logspace algorithm $A_s$ for LHOM$_s$($H$), as follows.

The algorithm $A_1$ simply checks if the mapping specified by the lists is a valid list homomorphism. That is, first we check for each vertex $v \in V(G)$ if $L(v) = \emptyset$. If any list is empty, there can be no list homomorphism and $A_1$ rejects. Otherwise we set up two counters to go through all pairs $(u,v) \in V(G)^2$. In each case, we check if $uv \in E(G)$, and if so, we check if $ab \in E(H)$, where $a$ and $b$ are the unique vertices in $L(u)$ and $L(v)$, respectively. If $ab$ is not an arc, $A_1$ rejects.

We assume next that $A_{s-1}$ is a logspace algorithm for LHOM$_{s-1}$($H$), and we show how to construct a logspace algorithm $A_s$ for LHOM$_s$($H$). Since $A_h$ is an algorithm for LHOM$(H)$, where $h = |V(H)|$, this will establish the algorithmic claim of Theorem~\ref{main_thm}. For the rest of this section, we focus on how to obtain $A_s$ from $A_{s-1}$.

Recall that a \emph{logspace transducer} is a Turing machine with a read-only input tape, a write-only output tape, and a worktape which can contain at most $O(\log n)$ symbols at any time. If $A$ and $B$ are two logspace transducers, we denote by $A \rightsquigarrow B$ the algorithm that first runs $A$, then feeds the output of $A$ to $B$, and eventually outputs the output of $B$. It is standard (see, e.g., \cite{sipser}) that $A \rightsquigarrow B$ can be implemented as a logspace transducer.

Assume that for each $k = 1, 2, \dots, m$ (recall that $m = |V(H^+)|$), there is a logspace transducer $T(a_k, b_k)$ such that on input $(G,L)$, where for each $v \in V(G)$ the list $L(v)$ is $k$-good, $T(a_k, b_k)$ outputs $(G,L')$ such that for each $v \in V(G)$, $L'(v)$ is $(k-1)$-good, and $G$ admits an $L$-homomorphism to $H$ if and only if $G$ admits an $L'$-homomorphism to $H$. Then notice that combining these transducers as
\[
T(a_m, b_m) \rightsquigarrow T(a_{m-1}, b_{m-1}) \rightsquigarrow \cdots \rightsquigarrow T(a_1, b_1) \rightsquigarrow A_1
\]
is in fact a logspace algorithm $A_s$ that correctly decides LHOM$_s$($H$). Hence, showing how to use $A_{s-1}$ to construct a transducer $T(a_k,b_k)$ for each $k = 1, 2, \dots, m$ will establish our claim. To simplify notation, we set $a = a_k$ and $b = b_k$.

\subsection{The Transducer $T(a,b)$}

Note that in what follows, we will implicitly use Reingold's algorithm \cite{reingold} on multiple occasions to decide whether two vertices are connected in the underlying undirected graph of certain digraphs. We summarize the desired properties of the instance $(G,L')$ returned by $T(a,b)$ on the input $(G,L)$:
\begin{itemize}
\item
each $L'(x) \subseteq L(x)$,
\item
no $L'(x)$ contains both $a$ and $b$, and
\item
$G$ admits an $L$-homomorphism to $H$ if and only if it admits an $L'$-homomorphism to $H$.
\end{itemize}
The second item means that $T(a,b)$ must remove at least one of $a$ or $b$ from each $L(x)$ that contains both, to obtain $L'$, that is, it makes $L'$ $(k-1)$-good.

We begin with defining the key notion of a {\em triple digraph} $Tr(G,L)$. The vertices of this digraph are triples $(y,c,d)$ where $y \in V(G),$ and $c, d \in L(y)$. Arcs of $Tr(G,L)$ are $(y,c,d)(y',c',d')$ such that $yy'$ is an arc of $G$, $cc'$ and $dd'$ are arcs of $H$, and $cd', dc'$ are not arcs of $H$.

The \emph{$ab$-test} is a logspace algorithm that will be used as a subroutine by the transducer $T(a,b)$. The \emph{$ab$-test} takes as input the triple digraph $Tr(G,L)$ and a vertex $(x,a,b)$ of this digraph, where we insist that the second and the third entries of the triple are $a$ and $b$, respectively. Since $Tr(G,L)$ will always be clear from the context, running the $ab$-test with input $(x,a,b)$ using $Tr(G,L)$ is simply denoted by $ab$-test$(x)$. The algorithm $ab$-test begins by constructing a sub-digraph $G''$ of $G$. Let $C(x,a,b)$ denote the weakly connected component of $(x,a,b)$ in $Tr(G,L)$. (Note that $C(x,a,b)$ can be constructed in logspace.) The digraph $G''$ consist of all vertices $y$ of $G$ such that $(y,c,d)$ is in $C(x,a,b)$ for some $c, d$. The arc $yy'$ belongs to $G''$ if there is an arc $(y,c,d)(y',c',d')$ in $C(x,a,b)$ for some $c, d, c', d'$. We define a new $L''$ by setting $L''(x)=a$, then defining $L''(y)$ for other $y \in V(G'')$ to consist of all $c \in L(y)$ for which

\begin{enumerate}
\item
there exists a $d \in V(H)$ such that $(y,c,d) \in C(x,a,b)$,
\item
there does not exist an $e \in V(H)$ such that $(y,e,c) \in C(x,a,b)$, and
\item
for every $z \not\in V(G'')$, if $zy$ is a forward (respectively backward) edge of $G$, then there exists a $t \in L(z)$ such that $tc$ is a forward (respectively backward) edge of $H$.
\end{enumerate}

Note at this point that it is not difficult to construct a logspace transducer that given $(G,L)$ outputs $Tr(G,L)$, and another logspace transducer that given $Tr(G,L)$ outputs $(G'',L'')$.

To finish the construction of the $ab$-test, $A_{s-1}$ is simulated with input $(G'',L'')$, and the output of the $ab$-test is the output of $A_{s-1}(G'',L'')$. Thus the $ab$-test on $x$ {\em succeeds} if there is an $L''$-homomorphism of $G''$ to $H$ taking $x$ to $a$. As we will see in the next section, for any $v \in V(G'')$, $|L''(v)| \leq s - 1$, so this last step is justified.

We are ready to define the core of the overall algorithm, the transducer $T(a,b)$. Before giving the formal definition below, we give a high-level description of $T(a,b)$, and also explain why precisely one of $a$ or $b$ is removed from the list of each vertex. Note that we need to handle only the \emph{relevant} vertices $x \in V(G)$, i.e., those vertices whose list contains both $a$ and $b$. Let $x_1,x_2,\dots,x_n$ be a list of the relevant vertices. The transducer $T(a,b)$ begins with determining if $x_1$ passes the $ab$-test. If so, then the \emph{group} of $x_1$ is defined to be the set of those vertices $y$ for which there exists a $c$ such that $(x_1,a,b)$ and $(y,c,a)$ are weakly connected in $Tr(G,L)$. The vertex $x_1$ is called the \emph{representative} of the group. $T(a,b)$ removes $b$ from $L(x_1)$, and $a$ from the list of every other vertex in this group. On the other hand, if $x_1$ does not pass the $ab$-test, then the only element of the group of $x_1$ is $x_1$ itself, and $T(a,b)$ removes $a$ from $L(x_1)$.

The transducer $T(a,b)$ finds the next vertex $x_i$ in the list $x_1,x_2,\dots,x_n$ which is not yet in the group of some $x_j$, $j < i$. Clearly, a vertex $x_i$ is in a previous group if and only if there is a $j < i$ such that the $ab$-test succeeds for $x_j$, and there is a $c$ such that $(x_j,a,b)$ and $(x_i,c,a)$ are weakly connected in $Tr(G,L)$. This can be easily determined by cycling through all $x_j$, $j < i$. Once the next $x_i$ is found, $T(a,b)$ defines the group of $x_i$ as for $x_1$. Of course, vertices that are already in some group are not placed again in the group of $x_i$. Group membership can be tested as before.

The formal description of $T(a,b)$ is given below. To facilitate exposition, the relevant vertices are specified in the input of $T(a,b)$ (we could also compute the relevant vertices from $(G,L)$ inside $T(a,b)$). Also note that since the vertices are written down on the input tape in some fixed order, we can traverse the relevant vertices according to the same fixed order whenever we wish to do so.
\begin{algorithm}[htb]
\renewcommand\thealgorithm{}
\caption{$\mathbf{T(a,b)}$}
\begin{algorithmic}[1]
\Require{a digraph with lists $(G,L)$, where the lists $L$ are $k$-good, vertices $x_1,\dots,x_n \in V(G)$ (in some fixed order) such that $a,b \in L(x_i)$, for each $i = 1,2, \dots, n$.}
\Ensure{a digraph with lists $(G,L')$, such that the lists $L'$ are $(k-1)$-good, and $G$ has an $L'$-homomorphism to $H$ if and only if $G$ has an $L$-homomorphism $H$.}
\Statex{(\emph{Note that the $ab$-test below is always with respect to $Tr(G,L)$.})}
\For{$i = 1 \to n$}
\If{for all $j < i$, it is not the case that for some $c$, $(x_i,c,a)$ is
\Statex{\pushcode[1] \qquad weakly connected to $(x_j,a,b)$ in $Tr(G,L)$ where \underline{$ab$-test}$(x_{j})$ succeeds}}\label{not_in_previous_group}
\If {\underline{$ab$-test}$(x_i)$ succeeds}\label{line3}
\State{\emph{Write down $L(x_i) \setminus b$ on the output tape}}\label{first_removal}
\For{$\ell = i+1 \to n$}
\If{for some $c$
\Statex{\pushcode[4] \qquad $(x_\ell,c,a)$ is weakly connected to $(x_i,a,b)$ in $Tr(G,L)$, and}
\Statex{\pushcode[4] \qquad $(x_\ell,c,a)$ is not weakly connected to any $(x_j,a,b)$, $j < i$, such that \underline{$ab$-test}$(x_j)$}
\Statex{\pushcode[4] \qquad  succeeds}}\label{other_group_elements_test}
\State\emph{Write down $L(x_{\ell}) \setminus a$ on the output tape}\label{second_removal}
\EndIf
\EndFor
\Else{ \emph{Write down $L(x_i) \setminus a$ on the output tape}}\label{third_removal}
\EndIf
\EndIf
\EndFor
\end{algorithmic}
\addtocounter{algorithm}{-1}
\end{algorithm}

It is easy to check that $T(a,b)$ can be implemented to have logarithmic space complexity: it uses a constant number of counters, uses the logspace subroutine $ab$-test, and also uses Reingold's logspace algorithm to test different (undirected) reachability conditions.

Before moving on to prove the correctness of the transducer $T(a,b)$, we link the formal definition to our high level description before. Based on the execution of $T(a,b)$, we partition the relevant vertices into \emph{groups}, and also define a \emph{representative} element for each group. We create a new group with representative $x_i$ whenever the condition in line 2 of $T(a,b)$ are satisfied. If the $ab$-test in line~\ref{line3} succeeds, then $b$ is removed from the list of $x_i$, and we also add all those elements $x_\ell$ ($\ell$ is defined in the loop in line 5) to the group of $x_i$ for which the conditions in line 6 are satisfied. Furthermore, $a$ is removed from the list of these additional vertices. But if the test in line~\ref{line3} fails, then the group consists only of the representative $x_i$, and $a$ is removed from its list. We say that a representative $x_i$ \emph{precedes} a representative $x_j$ if $i < j$, where the order is given in the input of $T(a,b)$. By the comments in the high level description, we can conclude Lemma~\ref{partition}.

\begin{lemma}\label{partition}
  The transducer $T(a,b)$ removes precisely one of $a$ or $b$ from the list of each relevant vertex.
\end{lemma}
\section{Correctness of $T(a,b)$} \label{sect-correct}

\subsection{Auxiliary Lemmas}

Let $p_1, p_2, \dots, p_m$ be the ordering of $V(H^+)$ described above, and recall that $p_k=(a,b)$. Let the strong component of $p_k$ be $C_{k'}$.

We introduce some notation. If $P$ and $Q$ are two walks such that the last vertex of $P$ is the same as the first vertex of $Q$, then $PQ$ denotes the walk obtained by first traversing $P$ and then $Q$. If $P$ is a walk of the form $a, a_1, a_2, \dots, a'$, then $P_j$ denotes the initial portion of this walk from $a$ to $a_j$. If $Q = b, b_1, b_2, \dots, b'$ is another walk congruent to $P$, then we note that if $P$ and $Q$ avoid each other, then $a \neq b$, $a_i \neq b_i$ for each $i$, and $a' \neq b'$.

\begin{lemma}\label{new}
Suppose $P = a, a_1, a_2, \dots, a', Q = b, b_1, b_2, \dots, b',$ and $R = b, c_1, c_2, \dots, c'$, are three congruent walks in $H$, such that, for every subscript $i$, the unordered pair $b_i,c_i$ is $k$-allowed.

If $P$ and $Q$ avoid each other, then $P$ and $R$ also avoid each other.
\end{lemma}

{\bf Proof:} For contradiction, assume $P$ and $R$ do not avoid each other; then there is a first faithful edge between $P$ and $R$. It could be some edge $c_ia_{i+1}$, or some edge $a_jc_{j+1}$ . Assume first that it is $c_ia_{i+1}$. Since it is the first faithful edge between $P$ and $R$, the walk $P_i$ avoids the walk $R_i$, and hence it is easy to see that the walk $R_i, a_{i+1}$ protects $Q_{i+1}$ from $P_{i+1}$. As $P_{i+1}, Q_{i+1}$ avoid each other, this is an extended $N$, and by Lemma \ref{helpful}, there is a circular $N$ in $H$, a contradiction. Therefore the first faithful edge is some $a_jc_{j+1}$. Note that the pair $(a_j,b_j)$ also belongs to the strong component $C_{k'}$ as the pair $p_k=(a,b)$ since $P$ and $Q$ avoid each other. Now we note that there is no faithful edge $b_jc_{j+1}$: suppose such an edge was in $H$, and assume without loss of generality that it is a forward edge, $b_jc_{j+1}$. This, together with the forward edges $a_jc_{j+1}$ and $b_jb_{j+1}$, and the fact that $a_jb_{j+1}$ is not a forward edge, means that the pair $(c_{j+1},b_{j+1})$ is in a strong component $C_{j'}$ with $j' > k'$, by the property $(*)$. This would mean that $(c_{j+1},b_{j+1})$ is some $p_s$ with $s > k$, contradicting our assumption that the unordered pair $c_{j+1},b_{j+1}$ is $k$-allowed. Thus there is no faithful edge $b_jc_{j+1}$, and the walks $Q_{j+1}$ and $P_j, c_{j+1}$ avoid each other. By the minimality of the subscript $j$, we again deduce that $R_{j+1}$ protects $Q_{j+1}$ from $P_j, c_{j+1}$, and obtain as above an extended, and then a circular, $N$, a contradiction.
\qed

A similar result applies if $R$ starts in $a$ instead of $b$.

\begin{lemma}\label{newer} Let $k = 1, 2, \dots, m$ be fixed.

Suppose $P = a, a_1, a_2, \dots, a', Q = b, b_1, b_2, \dots, b',$ and $R = a, c_1, c_2, \dots, c'$, are three congruent walks in $H$, such that, for every subscript $i$, the unordered pair $b_i,c_i$ is $k$-allowed.

If $P$ and $Q$ avoid each other, then $R$ and $Q$ also avoid each other.
\end{lemma}

{\bf Proof:}
For contradiction, assume $Q$ and $R$ do not avoid each other; then there is a first faithful edge (forward or backward) between $Q$ and $R$. It could be some edge $c_ib_{i+1}$, or some edge $b_jc_{j+1}$ (note that it could be both if $i=j$). Assume first that $c_ib_{i+1}$ is such a first faithful edge. Since it is a first faithful edge between $Q$ and $R$, the walk $Q_i$ avoids the walk $R_i$, and hence it is easy to see that the walk $R_i(c_ib_{i+1})$ protects $P_{i+1}$ from $Q_{i+1}$. As $P_{i+1}, Q_{i+1}$ avoid each other, this is an extended $N$, and by Lemma \ref{helpful}, there is a circular $N$ in $H$, a contradiction. Thus assume that the first faithful edge is some $b_jc_{j+1}$ (and $c_jb_{j+1}$ is not faithful). Note that the fact that this is a first faithful edge implies that the walks $Q_j$ and $R_j$ avoid each other, and so the pair $(c_j,b_j)$ also belongs to the strong component $C_{k'}$, as the pair $p_k=(a,b)$. Now note that $(c_j,b_j)(c_{j+1},b_{j+1})$ is a single arc of $H^+$ and hence by $(*)$ the pair $(c_{j+1},b_{j+1})$ is in a strong component $H^+$, which is different but reachable from the strong component containing $(c_j,b_j)$, a contradiction.
\qed

Note that the lemma implies that $b' \neq c'$.

\begin{cor}\label{newest} Let $k = 1, 2, \dots, m$ be fixed.

Suppose $P = c, a_1, a_2, \dots, a, Q = a, b_1, b_2, \dots, b$, and $R = a, c_1, c_2, \dots, d$, are three congruent walks in $H$, such that, for every subscript $i$, the unordered pairs $a_i,b_i$ and $b_i,c_i$ are $k$-allowed.

If $P$ and $Q$ avoid each other, then $P$ and $R$ also avoid each other.
\end{cor}

{\bf Proof:}
For contradiction, assume $P$ and $R$ do not avoid each other. If a first faithful edge is some $c_ia_{i+1}$, then the three walks $P' = P^{-1}, Q' = Q^{-1}$, and $R' = (P^{-1})_{i+1}(a_{i+1}c_i)(R_i)^{-1}$ contradict Lemma \ref{newer}, because the last two end in the same vertex. (Note that $R'$ begins at the end of $P$, follows the reversal of $P$ until $a_{i+1}$, then uses the edge $c_ia_{i+1}$ to go from $a_{i+1}$ to $c_i$, and then follows the reversal of $R$ from $c_i$ to $c_1$ and $a$.) On the other hand, if a first faithful edge is some $a_jc_{j+1}$, then we obtain walks contradicting Lemma \ref{newer} as follows: $P' = (P^{-1})_j(a_ja_{j+1})(P_{j+1})^{-1}, Q' = (Q^{-1})_j(b_jb_{j+1})(Q_{j+1})^{-1}, R' = (P^{-1})_j(a_jc_{j+1})(R_{j+1})^{-1}$.
\qed

We state the following useful fact.

\begin{cor}\label{spravne}
Suppose $(x',a',b')$ is reachable from $(x,a,b)$ in $Tr(G,L)$.

Then any $L$-homomorphism $f$ of $G$ to $H$ with $f(x)=a$ must have $f(x') \neq b'$.
\end{cor}

{\bf Proof:} Consider a walk from $(x,a,b)$ to $(x',a',b')$ in $Tr(G,L)$; it yields congruent walks $S, P,$ and $Q$ from $x$ to $x'$ in $G$, from $a$ to $a'$ in $H$, and from $b$ to $b'$ in $H$, respectively. Assume for contradiction that some $L$-homomorphism $f$ of $G$ to $H$ has $f(x)=a$ and $f(x')=b'$. Let $R$ be the image of the walk $S$ under $f$, from $a$ to $b'$. Now Lemma \ref{newer} implies that $R$ and $Q$ avoid each other, contradicting the fact that $R$ and $Q$ have the same last vertex. The lemma applies, because, for each $i$, the vertex $b_i$ is in $L(x_i)$ from the definition of $Tr(G,L)$, and the vertex $c_i$ is in $L(x_i)$ because $c_i=f(x_i)$ and $f$ is an $L$-homomorphism; thus the unordered pair $c_i,b_i$ is $k$-allowed.
\qed

\subsection{Analysis of $T(a,b)$}

Recall that the last step of the $ab$-test was calling the algorithm $A_{s-1}$ with input $(G'', L'')$. We now justify this step.

\begin{lemma}\label{extra}
For each vertex $y$ of $G''$ we have $|L''(y)| \leq |L(y)| - 1$.
\end{lemma}

{\bf Proof:}
This is clear for the vertex $x$, as $L(x)$ loses $b$. We note that any other $y$ in $G''$ yields a walk $Y$ from $x$ to $y$ in $G''$, and two corresponding walks in $H$, say $P$, from $a$ to some $c$, and $Q$, from $b$ to some $d$, that avoid each other. Note that $d \in L(y)$; however $d \not\in L''(y)$ since $(y,c,d)$ is in $C(x,a,b)$, i.e., is reachable from $(x,a,b)$ in $Tr(G,L)$.
\qed

As noted in Lemma~\ref{partition}, for any vertex of $G$ whose list contains both $a$ and $b$, $T(a,b)$ removes $a$ or $b$. Therefore the lists $L'$ returned by $T(a,b)$ are $(k-1)$-good. It remains to show that $G$ has an $L'$ homomorphism to $H$ if and only if $G$ has an $L$-homomorphism to $H$. Whether $a$ or $b$ was removed from the list of a given vertex depends on the outcome of the $ab$-test in line~\ref{line3}:
\begin{enumerate}
  \item If $ab$-test$(x)$ failed, then $a$ was removed from $L(x)$. This removal is justified by Lemma~\ref{ab-test_fails}.\label{first_type}
  \item If $ab$-test$(x)$ succeeded, then $b$ was removed from $L(x)$. Furthermore, $a$ was removed from $L(y)$ for any $y$ in the group of $x$. The role of Lemmas~{\ref{above}--\ref{combine_homs}} is to justify these removals.\label{second_type}
\end{enumerate}
Corollary~\ref{final} uses these lemmas to conclude the correctness of $T(a,b)$.

\begin{lemma}\label{ab-test_fails}
If there is an $L$-homomorphism $h$ from $G$ to $H$ such that $h(x)=a$, then the restriction of $h$ to $G''$ is an $L''$-homomorphism of $G''$ to $H$, and therefore $ab$-test$(x)$ in line~\ref{line3} of $T(a,b)$ succeeds.
\end{lemma}

{\bf Proof:}
We show that $h(y) \in L''(y)$ for any $y$ in $G''$.
As above, the fact that $y$ is in $G''$ implies that there is a path $Y$ from $x$ to $y$ in $G''$, and walks, $P$ from $a$ to some $a'$, and $Q$, from $b$ to some $b'$, that avoid each other. Let $R$ again denote the homomorphic image of $Y$ under $h$, i.e., $R$ is a walk from $h(x)=a$ to $h(y)$. By Lemma \ref{newer} $Q$ and $R$ also avoid each other. We now conclude that $h(y) \in L''(y)$, because no triple$(y,d,h(y))$ can be reached from $(x,a,b)$ in $Tr(G,L)$ by Corollary \ref{spravne}.
\qed

We move on to show that if there exists an $L$-homomorphism from $G$ to $H$, then there exists a homomorphism from $G$ to $H$ that respects the reduced lists of the vertices described in (\ref{second_type}). In Lemma~\ref{above}, we focus on constructing an $L$-homomorphism from $G$ to $H$ that respects the reduced lists of vertices of a given group. This key lemma provides such a homomorphism for \emph{each} group. Lemma~\ref{pairwise} combines two such homomorphisms to obtain an $L$-homomorphism from $G$ to $H$ that respects the reduced lists of each vertex in the two groups. Finally, Lemma~\ref{combine_homs} shows how to combine all the homomorphisms provided by Lemma~\ref{above} to obtain a homomorphism that respects the reduced lists of all vertices in all groups. Lemma~\ref{pairwise} is invoked in the induction step of Lemma~\ref{combine_homs}.

\begin{lemma}\label{above}
Let $\tilde{x}$ be a vertex in line~\ref{line3} of $T(a,b)$ for which the $ab$-test succeeded. If there exists an $L$-homomorphism $h$ of $G$ to $H$ such that $h(\tilde{x})=b$, then there exists an $L$-homomorphism $f$ of $G$ to $H$ such that $f(\tilde{x})=a$ and $f(y) \neq a$ for all other vertices $y$ in the group of $\tilde{x}$.

Similarly, if $\tilde{x}$ is a vertex for which the $ab$-test in line~\ref{line3} of $T(a,b)$ succeeded, and if there exists an $L$-homomorphism $h$ of $G$ to $H$ such that $h(z)=a$ for some $z$ in the group of $\tilde{x}$, then there exists an $L$-homomorphism $f$ of $G$ to $H$ such that $f(\tilde{x})=a$ and $f(y) \neq a$ for all other vertices $y$ in the group of $\tilde{x}$.
\end{lemma}

{\bf Proof:}
Let $h$ be a homomorphism of $G$ to $H$ as described above. Since the $ab$-test$(\tilde{x})$ succeeded, there exists an $L''$-homomorphism $g$ of $G''$ to $H$. Recall that $g(\tilde{x}) = a$. We define a mapping $f : V(G) \to V(H)$ as follows: $f(v)=g(v)$ if $v \in V(G'')$, and $f(v)=h(v)$ if $v \not\in V(G'')$. We claim that $f$ is a homomorphism of $G$ to $H$; since the lists $L''$ are subsets of the lists $L$, the mapping $f$ clearly respects the lists $L$, and we have $f(\tilde{x})=g(\tilde{x})=a$. Moreover, $f(y) \neq a$ for the other vertices in the group of $\tilde{x}$, by Corollary \ref{spravne}.

To prove the claim, assume that $yz$ is an arc of $G$ but $f(y)f(z)$ is not an arc of $H$. Since $g$ and $h$ are homomorphisms, this can only happen if one of $y, z$ is in $G''$ and the other one is not. Without loss of generality, we assume that $y \in V(G''), z \not\in V(G'')$. Thus $f(y)=g(y)$ and $f(z)=h(z)$. Assume $g(y)=a'$. The fact that $a'$ is in $L''(y)$ implies that there is a walk $Y$ from $\tilde{x}$ to $y$ in $G''$, and two corresponding walks $P = a, a_1, a_2, \dots, a'$ and $Q = b, b_1, b_2, \dots, b'$ in $H$ that avoid each other. Note that $P$ and $Q$ are both congruent to $Y$. We now introduce a third congruent walk, $R$, which is the homomorphic image of $Y$ under $h$, say, $h(\tilde{x})=b, c_1, c_2, \dots, h(y)$.  By Lemma \ref{new} $P$ and $R$ also avoid each other. (To see that the lemma applies, note again that $b_i, c_i$ all belong to the list of the $i$-th vertex of $Y$.) Thus the pair $(a',h(y))$ is in $C_{k'}$. Now we observe that $a'$ is in $L''(y)$ and therefore $a'$ satisfies property (3) in the description of the $ab$-test, and hence $a'd'$ is an arc of $H$ for some $d' \in L(z)$. Recall we are assuming that $f(y)f(z)=a'h(z)$ is not an arc of $H$. Clearly, $h(y)h(z)$ is an arc of $H$, so the pair $(d',h(z))$ is either in $C_{k'}$, if $h(y)d'$ is not an arc of $H$, or in a different strong component $C'$ reachable from $C_{k'}$, if $h(y)d'$ is an arc of $H$. Both cases contradict our assumptions. The second part of the lemma can be proved in a similar way after an application of Corollary~\ref{newest}.
\qed

Next, we take a closer look at two distinct representative vertices $\tilde{x}$ and $\tilde{x}'$ for which the $ab$-test succeeded. In particular, this means that there are $L$-homomorphisms $h, h'$ such that $h(\tilde{x})=h'(\tilde{x}')=a$. Note that $\tilde{x}, \tilde{x}'$ are not in the same group.

\begin{lemma}\label{pairwise}
Suppose $\tilde{x}$ and $\tilde{x}'$ are distinct representatives for which the $ab$-test in line~\ref{line3} of $T(a,b)$ succeeded, and assume that $\tilde{x}$ precedes $\tilde{x}'$. If $h, h'$ are two $L$-homomorphisms of $G$ to $H$ with $h(\tilde{x})=h'(\tilde{x}')=a$, then either
\begin{enumerate}
\item
$h'(\tilde{x}) \neq b$ and $h'(y) \neq a$ for all other vertices $y$ in the group of $\tilde{x}$ and the group of $\tilde{x}'$, or\label{original_hom}
\item
there exists another $L$-homomorphism $h''$ of $G$ to $H$ such that $h''(\tilde{x}) = h''(\tilde{x}') = a$, and $h''(y) \neq a$ for all $y$ in the group of $\tilde{x}$ and the group of $\tilde{x}'$; moreover, the value $h''(z)$ equals $h(z)$ or $h'(z)$ for all vertices $z$ in $G$.\label{h''_hom}
\end{enumerate}
\end{lemma}
{\bf Proof:}
If $h'(\tilde{x}) \neq b$ and $h'(y) \neq a$ for all other vertices $y$ in the group of $\tilde{x}$ then we have (\ref{original_hom}), because by Corollary \ref{spravne}, $h'(y) \neq a$ for all other vertices $y$ in the group of $\tilde{x}'$. Thus it remains to consider cases when $h'(\tilde{x})=b$, and when $h'(y)=a$ for some $y$ in the group of $\tilde{x}$.

We define $V^*$ to consist of those $z \in V(G)$ for which there exists a walk from $(\tilde{x},h(\tilde{x}),h'(\tilde{x}))$ to $(z,h(z),h'(z))$ in $Tr(G,L)$. We show that in both of the above cases $\tilde{x}'$ is not in $V^*$, and use this fact to define $h''$ in (\ref{h''_hom}).

\emph{Case 1:} $h'(\tilde{x})=b$. Since $\tilde{x}, \tilde{x}'$ are not in the same group, by definition, $(\tilde{x}',h(\tilde{x}'),a)$ is not reachable from $(\tilde{x},a,b)$ in $Tr(G,L)$. Since $(\tilde{x}',h(\tilde{x}'),h'(\tilde{x}')) = (\tilde{x}',h(\tilde{x}'),a)$  and $(\tilde{x},h(\tilde{x}),h'(\tilde{x})) = (\tilde{x},a,b)$, it follows that $\tilde{x}'$ is not in $V^*$.

\emph{Case 2:} $h'(y)=a$ for some $y$ in the group of $\tilde{x}$, and $h'(\tilde{x}) \neq b$. Since $y$ is in the group of $\tilde{x}$, there is a walk from $(\tilde{x},a,b)$ to some $(y,c,a)$ in $Tr(G,L)$ (see the figure below), corresponding to three congruent walks $S, P, Q$ from $\tilde{x}$ to $y$ in $G$, and from $a$ to $c$ and $b$ to $a$ in $H$, respectively, where $P$ and $Q$ avoid each other. Let $R$ be a third walk from $h'(\tilde{x})$ to $h'(y) = a$, the image of $S$ under $h'$. By Corollary \ref{newest}, $P$ and $R$ avoid each other, yielding a walk from $(y,c,a)$ to $(\tilde{x},a,h'(\tilde{x}))$ in $Tr(G,L)$.

If $\tilde{x}'$ was in $V^*$, then there would be a walk from $(\tilde{x},h(\tilde{x}),h'(\tilde{x}))$ to $(\tilde{x}',h(\tilde{x}'),h'(\tilde{x}'))$, i.e., from $(\tilde{x},a,h'(\tilde{x}))$ to $(\tilde{x}',h(\tilde{x}'),a)$ in $Tr(G,L)$. Therefore we would obtain a walk from $(\tilde{x},a,b)$ to $(\tilde{x}',h(\tilde{x}'),a)$ in $Tr(G,L)$, contradicting that $\tilde{x}'$ and $\tilde{x}$ are in different groups.

\begin{center}
  \includegraphics{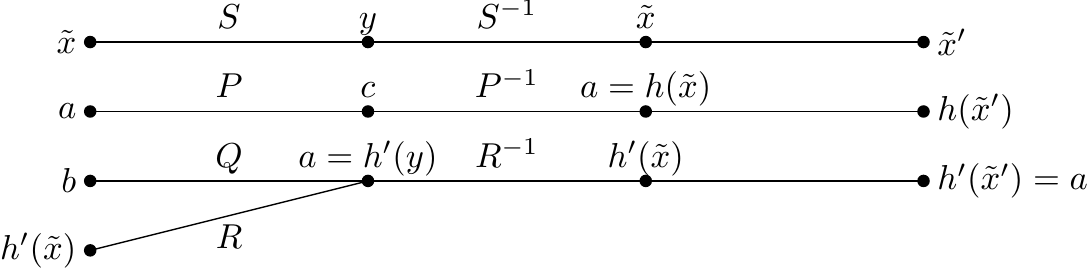}
\end{center}

It remains to show that if $\tilde{x}'$ is not in $V^*$, then $h''$ in (\ref{h''_hom}) exists. We define $h''$ to be $h$ on the vertices of $V^*$ and $h'$ on all other vertices. This mapping has $h''(\tilde{x})=h''(\tilde{x}')=a$, and is an $L$-homomorphism of $G$ to $H$. Indeed, assume that there is no arc between some $h(s)$ and $h'(t)$ where $st$ is an arc of $G$ and $s$ is in $V^*$ while $t$ is not. Since $h(s)$ has an arc with $h(t)$, and $h'(s)$ has an arc with $h'(t)$, $h'(s)$ must have an arc with $h(t)$, because otherwise $t$ is in $V^*$. But that would imply that $(h(t),h'(t))$ is not in $C_{k'}$ but in a component reachable from $C_{k'}$, yet $h(t), h'(t) \in L(t)$, contradicting our assumptions. (All of these arcs are in the same direction, forward or backward.) Since $h''(\tilde{x})=h''(\tilde{x}')=a$, we must have $h''(y) \neq a$ for all $y$ in the group of $\tilde{x}$ and the group of $\tilde{x}'$ by Corollary \ref{spravne}.
\qed

To complete proving the correctness of $T(a,b)$, let $\tilde{x}_1, \tilde{x}_2, \dots, \tilde{x}_m$ be the list of all the representatives for which the $ab$-test succeeded in line~\ref{line3} of $T(a,b)$, and for which there is an $L$-homomorphism from $G$ to $H$ that maps the representative to $b$. The elements of the list are ordered such that $\tilde{x}_i$ precedes $\tilde{x}_{i+1}$ for each $i = 1, \dots, m-1$. Note that this means (according to Lemma~\ref{above}) that for each $i = 1,2,\dots,m$, some $L$-homomorphism $h_i$ has $h_i(\tilde{x}_i)=a$ and $h_i(y) \neq a$ for all other $y$ in the group of $\tilde{x}_i$. We show that it was safe to remove $b$ from the list of all $\tilde{x}_i$ and $a$ from the lists of all other vertices in their groups. That is, we justify the removals outlined in (\ref{second_type}).

\begin{lemma}\label{combine_homs}
For each $i=1, 2, \dots, m$, let $h_i$ be an $L$-homomorphism from $G$ to $H$ such that $h_i(\tilde{x}_i)=a$, and $h_i(y) \neq a$ for any $y$ in the group of $\tilde{x}_i$. Then there exists an $L$-homomorphism $h$ from $G$ to $H$ such that for $i = 1, 2, \dots, m$, $h(\tilde{x}_i) \neq b$, and $h(y) \neq a$ for each $y$ in the group of $\tilde{x}_i$.
 \end{lemma}

{\bf Proof:} Let $S(j)$ be the statement such that for each $k = j, j+1, \dots, m$, there exists an $L$-homomorphisms $h$ from $G$ to $H$ such that
\begin{enumerate}
\item for each $i < j$, $h(\tilde{x}_i) \neq b$ and $h(y) \neq a$ for any $y$ that is in the group of $\tilde{x}_i$, and
\item $h(\tilde{x}_k) = a$ and $h(y) \neq a$ for any $y$ that is in the group of $\tilde{x}_k$.
\end{enumerate}

Observe that if we can establish $S(m)$ then the lemma follows. We induct on $j$. For the base case $S(1)$, notice that (1) does not apply, and (2) follows from the assumptions in the statement.

Assume that $S(j)$ holds, and fix an arbitrary $k \in \{j+1,\dots,m\}$. Since $S(j)$ holds, we have an $L$-homomorphism $h$ from $G$ to $H$ such that $h(\tilde{x}_i) \neq b$ for $i < j$ and $h(\tilde{x}_j) = a$.  Furthermore, $h$ does not map any of the non-representatives in the groups of $\tilde{x}_1,\dots,\tilde{x}_j$ to $a$. Similarly, there is another homomorphism $h'$ such that $h'(\tilde{x}_i) \neq b$ for $i < j$ and $h'(\tilde{x}_k) = a$. In addition, $h'$ does not map any of the non-representatives in the groups of $\tilde{x}_1,\dots,\tilde{x}_{j - 1}$ and $\tilde{x}_k$ to $a$. We apply Lemma~\ref{pairwise} with $h$, $h'$, $\tilde{x} = \tilde{x}_j$, and $\tilde{x}' = \tilde{x}_k$ (also observe that $\tilde{x}_j$ precedes $\tilde{x}_k$).

If the first case of Lemma~\ref{pairwise} applies, then $h'(\tilde{x}) = h'(\tilde{x}_j) \neq b$ and $h'(\tilde{x}_k) = a$. Furthermore, for any non-representative $y$ in the group of any of $\tilde{x}_1,\dots,\tilde{x}_j$ or $\tilde{x}_k$, $h(y) \neq a$. If the second case of Lemma~\ref{pairwise} applies, then there is an $L$-homomorphism $h''$ of $G$ to $H$ such that $h''(\tilde{x}) = h''(\tilde{x}_j) = a \neq b$, $h''(\tilde{x}_k) = a$. Using the fact that the value of $h''(z)$ equals the value of $h(z)$ or $h'(z)$ for any $z$, we can conclude that for any non-representative $y$ in the group of any of $\tilde{x}_1,\dots,\tilde{x}_j$ or $\tilde{x}_k$, $h(y) \neq a$.
\qed

\begin{cor}\label{final}
There is an $L$-homomorphism from $G$ to $H$ if and only if there is an $L'$-homomorphism from $G$ to $H$.
\end{cor}

{\bf Proof:}
If $G$ admits an $L'$-homomorphism to $H$, the same homomorphism is also an $L$-homomorphism from $G$ to $H$ because $L'(x) \subseteq L(x)$ for any $x$.

For the converse, observe that the homomorphisms $h_1,\dots,h_m$ required by Lemma~\ref{combine_homs} are provided by Lemma~\ref{above}, and thus Lemma~\ref{combine_homs} yields an $L$-homomorphism $h$ from $G$ to $H$. Notice that by the properties of $h$ specified in Lemma~\ref{combine_homs}, and by Lemma~\ref{ab-test_fails} (also see the paragraph after the lemma), $h$ is actually an $L'$-homomorphism from $G$ to $H$.

(Note that $f$ also respects the $L'$-lists of those representatives which do not appear in the list $\tilde{x}_1,\dots,\tilde{x}_m$, and the $L'$-lists of the elements in the groups of these representatives. The reason is that in the proofs of Lemmas~\ref{above},\ref{pairwise}, and \ref{combine_homs}, these vertices never get mapped to something outside the $L'$-lists.)
\qed

\section{Algebraic Results} \label{sect-nice-terms}

As mentioned in the introduction, the conjecture of Larose and Tesson predicts that if $H$ admits a so-called finite HM-chain of polymorphism, then LHOM($H$) should be solvable in logspace, and the absence of such a chain is already known to imply NL-hardness \cite{lartes}. We now define the concept an HM-chain (first defined in \cite{hagmit}). We will then prove that the digraphs that admit such a chain of polymorphisms are precisely those without a circular $N$ (Theorem \ref{thm51}.)

\begin{dfn} Let $H$ be a digraph. An operation $f : V(H)^m \rightarrow V(H)$
is a polymorphism of $H$ if $f(v_{11},v_{12},\dots,v_{1m})f(v_{21},v_{22},\dots,v_{2m}) \in E(H)$ whenever
$v_{11}v_{21},v_{12}v_{22},\dots,v_{1m}v_{2m} \in E(H)$.
\end{dfn}

\begin{dfn} \label{def-HM}
  A sequence $f_1, \dots, f_k$ of ternary operations is called a \emph{Hagemann-Mitschke chain of length $k$ (HM-chain)} if it satisfies the identities
\begin{itemize}
    \item $x = f_1(x,y,y)$
    \item $f_i(x,x,y) = f_{i+1}(x,y,y)$ for all $i=1, \dots, k-1$
    \item $f_k(x,x,y)=y.$
\end{itemize}
We say that $H$ {\em admits} an HM-chain $f_1,f_2,\dots,f_k$ if each $f_i$ is a polymorphism of $H$.
\end{dfn}

We prove the following main result of the section.

\begin{tm}\label{thm51}
Let $H$ be a digraph. Then $H$ does not admit an HM-chain of finite length if and only if $H$ contains a circular $N$.
\end{tm}

We begin with some definitions and observations. Let $(x,y)$ be a vertex of $H^+$, lying in a strong component $K$. We denote by
$\mu(x,y)$ the maximum number of vertices in a directed path ending at $K$, in the
condensation $C(H^+)$.

By Lemma \ref{N} and a nearby observation, $\mu(x,y)$ is one plus the maximum
number of single edges in a forward walk of $H^+$ that ends at $(x,y)$. In other words,
for any vertex $(x,y)$ there exist congruent walks $X, Y$ in $H$, ending at $x, y$
respectively, such that $X$ avoids $Y$, and $Y$ has $\mu(x,y) - 1$ faithful edges
to $X$.

We note for future use that it follows from the definitions that if $(x,y)(x',y')$
is an arc of $H^+$ (meaning in particular that $xy'$ is not an edge of $H$),
then $\mu(x,y) \leq \mu(x',y')$; and if $(x,y)(x',y')$ is a single arc of $H^+$
(meaning that $yx'$ is an edge of $H$), then $\mu(x,y) < \mu(x',y')$.

We also need the following simple observation.
\begin{obs}\label{same_strong}
  Let $(x,y)$ and $(x',y')$ be two vertices of $H^+$. If $\mu(x,y) = \mu(x',y')$ and there is directed walk from $(x,y)$ to $(x',y')$ in $H^+$, then $(x,y)$ and $(x',y')$ are in the same strong component of $H^+$.
\end{obs}

Theorem~\ref{thm51} follows from the two lemmas below.

\begin{lemma}\label{cN-noHM}
Let $H$ be a digraph. If $H$ has a circular $N$, then $H$ admits no HM-chain of any length.
\end{lemma}
\pf Suppose $H$ admits an HM-chain $f_1, f_2, \dots, f_k$. Let $X = x_0,x_1,\dots,x_n$, $Y = y_0,y_1,\dots,y_n$ and $Z = z_0,z_1,\dots,z_n$ be the walks that form a circular $N$, where indices are modulo $n$. That is, $X$ avoids $Y$, and $Z$ protects $Y$ from $X$. We claim that if $f_i(x_0,y_0,y_0) = x_0$, then $f_i(x_0,x_0,y_0) = x_0$, $0 \leq i \leq n$. This, together with $f_1(x_0,y_0,y_0) = x_0$, $f_k(x_0,x_0,y_0) = y_0$, and $f_i(x_0,x_0,y_0) = f_{i+1}(x_0,y_0,y_0)$ by definition, implies that $x_0 = y_0$, contradicting that $X$, $Y$ and $Z$ form a circular $N$.

Let $\ell$ be the smallest integer such that $x_{\ell}z_{\ell+1}$ is an edge. Observe that $f_i(x_0,y_0,y_0),f_i(x_1,z_1,y_1),$
$\dots,f_i(x_{\ell+1},z_{\ell+1},y_{\ell+1})$ is a walk from $x_0$ to $f_i(x_{\ell+1},z_{\ell+1},y_{\ell+1})$ (congruent to the sub-walk of $X$ from $x_0$ to $x_{\ell+1}$). Since $f_i$ is conservative, $X$ avoids $Y$, and the sub-walk of $X$ from $x_0$ to $x_\ell$ avoids the sub-walk of $Z$ from $z_0$ to $z_\ell$ by the choice of $\ell$, it must be that $f_i(x_{\ell+1},z_{\ell+1},y_{\ell+1}) \in \{x_{\ell+1},z_{\ell+1}\}$. Because $Z$ protects $Y$ from $X$, the sub-walk of $Z$ from $z_{\ell+1}$ to $z_n$ avoids the sub-walk of $Y$ from $y_{\ell+1}$ to $y_n$. It follows that the walk $f_i(x_{\ell+1},z_{\ell+1},y_{\ell+1})$, $f_i(x_{\ell+2},z_{\ell+2},y_{\ell+2}), \dots, f_i(x_n,z_n,y_n)$ must end at $x_n$, i.e., $x_n = f_i(x_n,z_n,y_n) = f_i(x_n,x_n,y_n) = f_i(x_0,x_0,y_0) = x_0$.

\qed

\begin{lemma}\label{noCN->HM}
Let $H$ be a digraph. If $H$ has no circular $N$, then $H$ admits an HM-chain of length $k$, where $k$ is the number of vertices in the longest directed path in $C(H^+)$.
\end{lemma}
\pf
Suppose that $C(H^+)$ has no directed path of length $k$; in that case,
$\mu(x,y) \leq k$ for all vertices $(x,y)$ of $H^+$. We define an HM-chain $f_1, f_2, \dots, f_k$ using the values $\mu$, and the following notion of an
{\em $i$-distinguisher}.

For vertices $a, b, c$ of $H$, we say that $a$ is an {\em $i$-distinguisher} of
$a, b, c$ if there exist vertices $x$ and $y$ with $\mu(x,y) \geq i$, and three
congruent walks, $X$ from $a$ to $x$, $Y$ from $b$ to $y$, and $Z$
from $c$ to $y$ such that both $Y$ and $Z$ avoid $X$. We shall write
$a = d_i(a,b,c)$ to mean that $a$ is an $i$-distinguisher of $a, b, c$.

We now define $f_i(a,b,c)$ as follows.

\begin{itemize}
\item[{\bf Case A}] Suppose $\mu(b,c) > i$

\begin{enumerate}
\item
if $\mu(a,b) < i$, then $f_i(a,b,c)=b$
\item
if $\mu(a,b) \geq i$, then $f_i(a,b,c)=a$
\end{enumerate}

\item[{\bf Case B}] Suppose $\mu(b,c) = i$

\begin{enumerate}
\item
if $\mu(a,c) < i$, or if $\mu(a,c) = i$ and $a \neq d_i(a,b,c)$, then $f_i(a,b,c)=c$
\item
if $\mu(a,c) > i$, or if $\mu(a,c) = i$ and $a = d_i(a,b,c)$, then $f_i(a,b,c)=a$
\end{enumerate}

\item[{\bf Case C}] Suppose $\mu(b,c) < i$

\begin{enumerate}
\item
if $\mu(a,c) < i$, then $f_i(a,b,c)=c$
\item
if $\mu(a,c) \geq i$, then $f_i(a,b,c)=a$
\end{enumerate}

\end{itemize}

The definition will be applied to all triples $a, b, c$, whether or not the vertices $a, b, c$
are distinct, with the convention that $\mu(x,x) = 0$. Thus from the first part of Case A we
obtain

\begin{equation*}
f_i(x,x,y) = x \text{ if } \mu(x,y) > i
\end{equation*}

\noindent and from the first parts of cases B (with $a \neq d_i(a,b,c)$) and C, we obtain

\begin{equation}\label{xxy=y}
f_i(x,x,y) = y \text{ if } \mu(x,y) \leq i.
\end{equation}

\noindent Similarly, Case C yields

\begin{equation}\label{two_cases}
  f_i(x,y,y) = y  \text{ if } \mu(x,y) < i \text{ and } f_i(x,y,y) = x \text{ if } \mu(x,y) \geq i.
\end{equation}

\begin{claim}
  The operations $f_1, f_2, \dots, f_k$ form an HM-chain.
\end{claim}
\pf
Indeed, since we always have
$\mu(x,y) \leq k$, we must have $f_k(x,x,y)=y$. Similarly we always have $\mu(x,y) \geq 1$,
hence $f_1(x,y,y) = x$. Moreover, if $f_i(x,x,y)=x$, then $i < \mu(x,y)$ and hence
$f_{i+1}(x,y,y)= x$, and if $f_i(x,x,y)=y$ then $\mu(x,y) \le i$ and hence $f_{i+1}(x,y,y)=y$.
\qed

\begin{claim}
The operations $f_1,f_2,\dots,f_k$ are polymorphisms of $H$.
\end{claim}
\pf
Let $aa',bb',cc'$ be forward (backward) edges in $H$, and assume for contradiction that $f_i(a,b,c)f_i(a',b',c')$ is \emph{not} an edge. We have to consider the following six cases: $aa',bb',cc'$ are forward (backward) edges and one of $ba'$, $bc'$ or $ca'$ is not a forward (backward) edge. We analyse only the cases when $aa',bb',cc'$ are all forward edges and one of $ba'$, $bc'$ or $ca'$ is not a forward edge. The analysis of the cases when all these edges are backward edges can be carried out in an identical way (the reader might find it useful to recall the definition of $H^+$).

{\bf Case 1} $f_i(a,b,c)=b, f_i(a',b',c')=a'$, and $ba'$ is not an edge. We argue first that $i > \mu(a,b) \geq \mu(a',b')$. Note that since $ba'$ is not an edge, $a \neq b$ and $a' \neq b'$. If $b \neq c$, then by the definition of $f_i(a,b,c)$, the only way $f_i(a,b,c)$ can be $b$ is if $\mu(a,b) < i$. If $b = c$, then (\ref{two_cases}) above implies that $\mu(a,b) < i$. Also observe that since $(a',b')(a,b)$ is an edge of $H^+$, we have that $i > \mu(a,b) \geq \mu(a',b')$ (the second inequality follows from the remarks before the statement of Theorem~\ref{thm51}).

We claim that $\mu(b',c') \leq i$. If $i > \mu(a',b')$ and $\mu(b',c') > i$, then $f_i(a',b',c')=b'$, contradicting that $a' \neq b'$. Hence $\mu(b',c') \leq i$.

We claim that $bc'$ is an edge. If $bc'$ is not an edge, then $\mu(b,c) \leq \mu(b',c') \leq i$. It follows that $f_i(a,b,c) \in \{a,c\}$. As noted earlier, $a \neq b$, so we must have that $f_i(a,b,c) = b = c$. But if $b = c$ then $bc'$ is an edge.

Since $bc'$ is an edge, $(a',c')(a,b)$ an edge of $H^+$, and therefore $i > \mu(a,b) \geq \mu(a',c')$. This together with $\mu(b',c') \leq i$ implies that $f_i(a',b',c')=c'$, and therefore $c' = a'$. Because $bc'$ is an edge, $ba'$ is an edge, a contradiction.

{\bf Case 2} $f_i(a,b,c)=b, f_i(a',b',c')=c'$, and $bc'$ is not an edge.

We argue that $\mu(b',c') > i$ and $\mu(b,c) > i$. Observe first that $(b,c)(b',c')$ is an edge of $H^+$ and hence $\mu(b',c') \geq \mu(b,c)$. To see that $\mu(b,c) > i$, assume that $\mu(b,c) \leq i$. Then by the definition of $f_i$, $f_i(a,b,c) \in \{a,c\}$, hence $b=a$ or $b=c$. Since $cc'$ is an edge and $bc'$ is not, $b \neq c$. If $a = b$, then (\ref{xxy=y}) above gives that $f_i(a,b,c) = c$, contradicting that $b \neq c$.

Now $\mu(b',c') > i$ implies that $f_i(a',b',c')$ is either
$a'$ or $b'$, and therefore $a'=c'$ or $b'=c'$. Note that $b'=c'$ is impossible, since $bc'$ is not an edge. Thus consider the case when $a'=c'$. Since $\mu(b',c') > i$ and $f_i(a',b',c')=c' = a'$, it follows from the definition of $f_i$ that $\mu(a',b') \geq i$. Since $a' = c'$ and we are assuming there that $bc'$ is not an edge, $(a',b')(a,b)$ is an edge of $H^+$. Therefore $\mu(a,b) \geq \mu(a',b') \geq i$. This together with $\mu(b,c) > i$ implies that$f_i(a,b,c)=a$, and therefore $a=b$. However, $a=b, a'=c'$ is impossible as $aa'$ is an edge and
$bc'$ is not.

{\bf Case 3} $f_i(a,b,c)=c, f_i(a',b',c')=a'$, and $ca'$ is not an edge.

Note first that $(a',c')(a,c)$ is an edge of $H^+$ and hence $\mu(a,c) \geq \mu(a',c')$.
Also, we must have $a' \neq c'$ and $a \neq c$ as $cc'$ and $aa'$ are edges.

If $f_i(a,b,c)=c$, then by the definition of $f_i$, we have three possible cases:
\begin{enumerate}
  \item $\mu(b,c) > i$, $\mu(a,b) < i$, $f_i(a,b,c) = b = c$ (Case A, (1))
  \item $\mu(b,c) \leq i$ and $\mu(a,c) < i$ (Case B, first half of (1), and Case C (1))
  \item or $\mu(b,c)=\mu(a,c)=i$ and $a \neq d_i(a,b,c)$ (Case B, second half of (1)).
\end{enumerate}
The first case is impossible since if $b = c$, then $\mu(b,c) = 0$ and $i \geq 1$. In the second case, $\mu(a',c') < i$, and hence $\mu(b',c') \geq i$ (else $f_i(a',b',c')=c'$). Now if $cb'$ is not an edge we have $\mu(b,c) \geq \mu(b',c')$ and hence $\mu(b',c') = i$, implying that
$f_i(a',b',c')=c'$, a contradiction. On the other hand, if $cb'$ is an edge, then $(a',b')(a,c)$ is an edge of
$H^+$ and hence $i > \mu(a,c) \geq \mu(a',b')$. If $\mu(b',c') \leq i$ then we have
$f_i(a',b',c')=c'$ because of $\mu(a',c') < i$, and if $\mu(b',c') > i$ then we have
$f_i(a',b',c')=b'$ because of $\mu(a',b') < i$. Since neither $c'$ nor $b'$ can be equal
to $a'$, these contradictions prove that the first situation cannot occur.

The rest of the proof is devoted to showing that situation (3) is also impossible. We keep in mind that $\mu(a,c) \geq \mu(a',c')$ and therefore $\mu(a',c') \leq i$.

{\bf We first assume that $ac'$ is an edge.}

Our goal is to show that $a$ is an $i$-distinguisher of $a,b,c$, and this will contradict the above assumptions. To do this, we show that $\mu(a',b') \geq i$, there are three congruent walks (in fact, just forward edges), $X = aa'$, $Y = bb'$, and $Z = cb'$ such that both $Y$ and $Z$ avoid $X$, that is, $ca'$ and $ba'$ are not forward edges.

We show that $\mu(b',c') > i$. Assume that $\mu(b',c') \leq i$. Then the assumption that $ac'$ is an edge implies that $\mu(a',c') < \mu(a,c) = i$, and it follows that $f_i(a',b',c')=c'$. This is impossible because $a' \neq c'$.

\textbf{$cb'$ is a forward edge:} If not, then $(b',c')(b,c)$ is an edge of
$H^+$ and hence $\mu(b',c') \leq \mu(b,c) = i$, contradicting that $\mu(b',c') > i$.

\textbf{$ba'$ is not an edge:} If it is, then $\mu(a',b') < \mu(b,c) = i$, and this, together with $\mu(b',c') > i$, gives that $f_i(a',b',c')=b'$. This implies that $a' = b'$, which is not possible because $cb'$ is an edge.

\textbf{$\mu(a',b') \geq i$:} otherwise the fact that $\mu(b',c') > i$ implies a contradiction (as in the previous paragraph).

{\bf We now assume that $ac'$ is not an edge.}

In this case, we show that $ab'$, $ba'$, $bc'$ and $cb'$ are not forward edges, and use this to obtain a contradiction. Observe that $(a,c)$ and $(a',c')$ are in the same strong component of $H^+$ and hence $\mu(a',c')=i$. In what follows, we will use this fact several times.

\textbf{$ba'$ is not an edge.} Assume that $ba'$ is an edge. Then $(a',c')(b,c)$ is an edge of $H^+$; it cannot be a single edge of $H^+$ since $\mu(a',c')=i$ and $\mu(b,c)=i$. Thus it must be a double edge of $H^+$ and hence $bc'$ is not an edge
of $H$. We show how to get a contradiction whether or not $cb'$ is an edge.

If $cb'$ is an edge, then $(a',b')(b,c)$ is a single edge of $H^+$ and hence $\mu(a',b') < i$; at the same time, $(b,c)(b',c')$ is a single edge of $H^+$ and hence $\mu(b',c') > i$, implying that $f_i(a',b',c')=b'$, contrary to our assumption. (We have $b' \neq a'$ as $cb'$ is an edge.)

If $cb'$ is not an edge, then $(a',c')(b,c)$ and $(b,c)(b',c')$ are double edges of $H^+$, and hence
$(a',c')$ and $(b',c')$ are in the same strong component of $H^+$. Therefore $\mu(a',c')=\mu(b',c')=i$, and since $f_i(a',b',c')=a' \neq c'$, the definition of $f_i$ implies that $a' = d_i(a',b',c')$.

Let $X'$, $Y'$, and $Z'$ be the walks from $a'$ to $x$, $b'$ to $y$, and $c'$ to $y$, respectively, witnessing that $a' = d_i(a',b',c')$. Then both $Y'$ and $Z'$ avoid $X'$, and $\mu(x,y) \geq i$. Set $X = X'^{-1}b$, $Y = Y'^{-1}b$ and $Z = Z'^{-1}c$. Then $X$, $Y$ and $Z$ form an extended $N$ from $x,y$ to $b,c$: $X$ avoids $Z$ since $ca'$ is not an edge, and $Y$ protects $Z$ from $X$ since $X'^{-1}$ avoids $Y'^{-1}$.

We claim further that $Z$ avoids $X$. Indeed, since there is a directed walk $W$ from $(x,y)$ to $(b,c)$ (using $X$, $Z$), $\mu(x,y) \ge i$ and $\mu(a,c)=i$, we conclude that $\mu(x,y)=i$. It follows from Observation~\ref{same_strong} that $(x,y)$, $(b,c)$, and $W$ are all in the same strong component of $H^+$. Then Observation~\ref{same_strong} implies that all edges of $W$ are double, and therefore $Z$ avoids $X$. We now apply Lemma~\ref{helpful} to obtain a circular $N$, a contradiction.

\textbf{$cb'$ is not an edge.} Assume that $cb'$ is an edge. Then $ab'$ cannot be an edge for the following reason. If $ab'$ is an edge then $(a,c)(b',c')$ is a single edge of $H^+$, and therefore $\mu(b',c') > i$. Moreover, $(a',b')(a,c)$ is also a single edge of $H^+$, so
$\mu(a',b') < i$. Hence, we have that $f_i(a',b',c')=b' \neq a'$, and this is a contradiction. So $ab'$ is not an edge. It follows that there is a double edge between $(a',b')$ and $(a,c)$ in $H^+$, so $i = \mu(a,c) = \mu(a',b')$. Using the walks
$X = aa'$, $Y = bb'$, and $Z = cb'$, and the fact that $ba'$ and $ca'$ are not edges (so $Y$ and $Z$ avoid $X$), we obtain that $a = d_i(a,b,c)$. This contradicts our assumptions.

\textbf{$bc'$ is not an edge.} Assume that $bc'$ is an edge. Then using the walks $X=aa'$, $Y=bc'$ and $Z=cc'$, the fact that $ba',ca'$ are not edges, and recalling that $\mu(a',c')=i$, we obtain again that $a = d_i(a,b,c)$.

\textbf{$ab'$ is not an edge.} Assume that $ab'$ is an edge. Since $bc'$ and $cb'$ are not edges, there is a double edge between $(b,c)$ and $(b',c')$ in $H^+$, and therefore $i = \mu(b,c) = \mu(b',c')$. Recalling that $(a',c')=i$, the definition of $f_i$ implies that $a' = d_i(a',b',c')$. This means that there exist $x, y$ and walks $X'$ from $a'$ to $x$, $Y'$ from $b'$ to $y$ and
$Z'$ from $c'$ to $y$ such that $\mu(x,y) \geq i$ and $Y'$ and $Z'$ avoid $X'$. Set $X=aX'$, $Y=aY'$, and $Z=cZ'$.
Similarly to the proof that $ba'$ is not an edge, we can verify that $X, Y, Z$ form an extended $N$ from $c,a$ to $y,x$, and that $X$ avoids $Z$. Under these conditions, Lemma~\ref{helpful} guarantees the existence of a circular $N$, which is a contradiction.

We now have $aa',bb',cc'$ as the only edges from $a,b,c$ to $a',b',c'$. In particular, $(b,c)$ and $(b',c')$ are in the same strong component of $H^+$, and therefore $i = \mu(b,c) = \mu(b',c')$. Recalling that $\mu(a'c') = i$, we obtain again that $a'=d_i(a', b',c')$. Since $ba', ca'$ are not edges, also $a=d_i(a,b,c)$, contrary to our assumption. This completes the proof.
\qed

We conclude the proof of Theorem~\ref{thm51}:

\textbf{Proof of Theorem~\ref{thm51}}
If $H$ contains a circular $N$, then we use Lemma~\ref{cN-noHM}. For the converse, assume that $H$ contains no circular $N$. Recalling that $C(H^+)$ is acyclic and therefore the longest directed path has bounded length, now we can apply Lemma~\ref{noCN->HM}.
\qed
\section{Additional Remarks} \label{sect-additional}

In this section, we summarize the known results about the complexity of LHOM$(H)$, and conclude with a polynomial time algorithm to decide if a digraph contains a circular $N$. For motivation and definitions related to first-order definability, the reader can consult \cite{lalota,adrien}. We begin by giving the definition of a DAT.

\begin{dfn}[\cite{soda}] \label{def-dat}
A \emph{digraph asteroidal triple} (DAT) is a triple of vertices $u,v,w$ together with six vertices $s(u),b(u),
s(v),b(v), s(w),b(w)$, which satisfy the following conditions:
\begin{enumerate}
  \item For each permutation $x,y,z$ of $u,v,w$, there exists a walk $P(x,s(x))$ from $x$ to $s(x)$ and two walks $P(y,b(x))$ (from $y$ to $b(x)$), and $P(z,b(x))$ (from $z$ to $b(x)$), congruent to $P(x,s(x))$, such that $P(x,s(x))$ avoids both $P(y,b(x))$ and $P(z,b(x))$.
  \item Each of the three pairs $(s(u),b(u))$, $(s(v),b(v))$, and $(s(w),b(w))$ is invertible.
\end{enumerate}
\end{dfn}

\begin{tm}[\cite{soda}]
  Let $H$ be a digraph. If $H$ contains a DAT, the problem LHOM$(H)$ is NP-complete. If $H$ is DAT-free, the problem LHOM($H$) is polynomial time solvable.
\end{tm}

As one would expect, the presence of a DAT implies the presence of a circular $N$.

\begin{prp}\label{DAT->cN}
  If a digraph $H$ contains a DAT, then it contains a circular $N$.
\end{prp}
\pf
Let $u,v,w$ be the triple of vertices in the DAT. It follows from Theorem 3.2 of \cite{soda} that all pairs $(s(u),b(u))$, $(s(v),b(v))$, and $(s(w),b(w))$ and their inverses belong to the same strong component of $H^+$. In particular, these pairs are invertible. This fact will be used in the proof.

Assume that $H$ contains a DAT. We break down the proof into two parts. First, we assume the non-existence of certain edges, and using this assumption, we show the presence of a circular $N$ in $H$. The second part of the proof assumes that at least one of the previous edges is present, and then this is used to construct a circular $N$ in a different way.

\begin{figure}[!htb]\label{makes_sense_fig}
\begin{center}
  \includegraphics{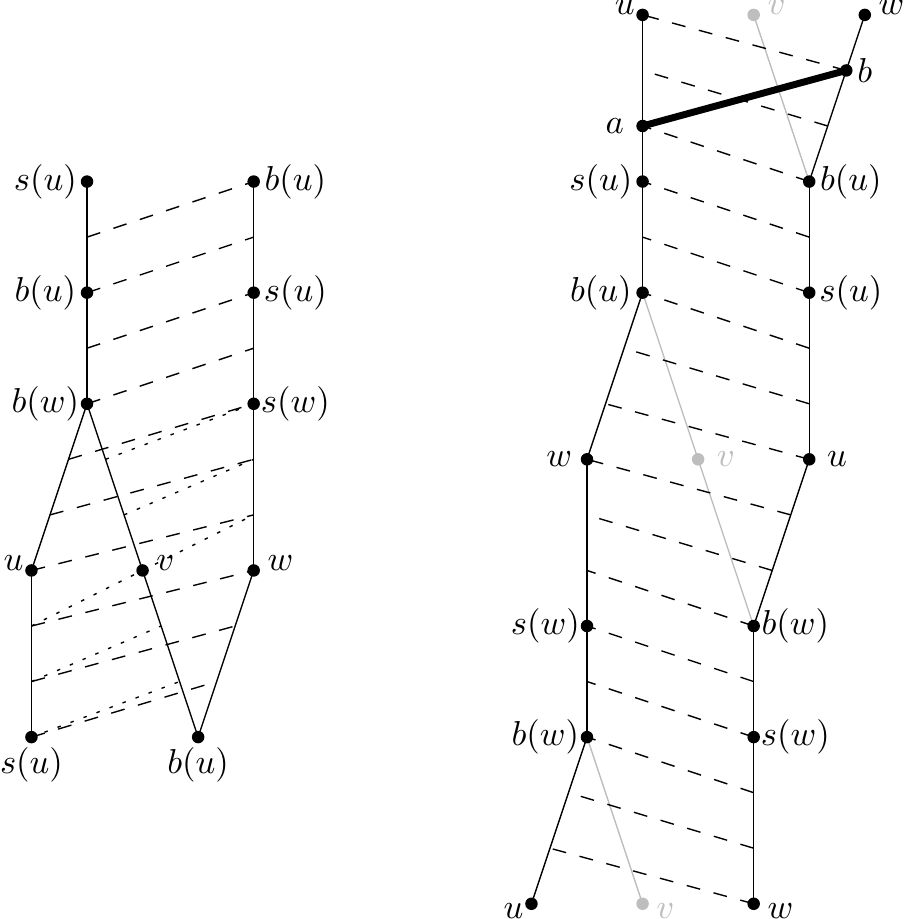}
\end{center}\caption{Circular $N$-s in the proof of Proposition~\ref{DAT->cN}. The dashed and dotted lines represent missing edges.}
\end{figure}

Let $u,v,w$ be a DAT. First we assume that $P(u,s(u))^{-1}$ avoids both $P(v,b(u))^{-1}$ and $P(w,b(u))^{-1}$, and both $P(u,b(w))$ and $P(v,b(w))$ avoid $P(w,s(w))$ (this case is illustrated on the left side of Figure~\ref{makes_sense_fig}; all walks in the figure go from bottom to top). Furthermore, observe that since $(b(w),s(w))$ and $(b(u),s(u))$ are in the same strong component of $H^+$, there are congruent walks $P(b(w),b(u))$ from $b(w)$ to $b(u)$ and $P(s(w),s(u))$ from $s(w)$ to $s(u)$ such that $P(b(w),b(u))$ avoids $P(s(w),s(u))$. Since $(b(u),s(u))$ is invertible, we have congruent walks $P(b(u),s(u))$ from $b(u)$ to $s(u)$ and $P(s(u),b(u))$ from $s(u)$ to $b(u)$, such that $P(b(u),s(u))$ avoids $P(s(u),b(u))$. Therefore the walks
\begin{align*}
    X &= P(u,s(u))^{-1} P(u,b(w)) P(b(w),b(u)) P(b(u),s(u)),\\
    Y &= P(w,b(u))^{-1} P(w,s(w)) P(s(w),s(u)) P(s(u),b(u)),\\
    Z &= P(v,b(u))^{-1} P(v,b(w)) P(b(w),b(u)) P(b(u),s(u))
\end{align*}
form a circular $N$.

For the second case, there is a faithful edge $ab$ either from $P(u,s(u))^{-1}$ to $P(w,b(u))^{-1}$, or from $P(u,s(u))^{-1}$ to $P(v,b(u))^{-1}$, or from $P(u,b(w))$ to $P(w,s(w))$, or from $P(v,b(w))$ to $P(w,s(w))$. We show how to find a circular $N$ in first case (see the right side of Figure~\ref{makes_sense_fig}), and the other three cases can be handled analogously. As in the first case, we can find a walk from $P(b(u),s(u))$ from $b(u)$ to $s(u)$ and a walk $P(s(u),b(u))$ from $s(u)$ to $b(u)$ such that the two walks are congruent and $P(s(u),b(u))$ avoids $P(b(u),s(u))$.

By the definition of a DAT, there are congruent walks $P(w,b(u))$ from $w$ to $b(u)$ and $P(u,s(u))$ from $u$ to $s(u)$ such that $P(u,s(u))$ avoids $P(w,b(u))$. This completes the top half of the right side of the figure.

Using similar arguments, there are congruent walks $P(w,s(w))^{-1}$ from $s(w)$ to $w$ and $P(u,b(w))^{-1}$ from $b(w)$ to $u$ such that $P(u,b(w))^{-1}$ avoids $P(w,s(w))^{-1}$. There are congruent walks $P(b(w),s(w))$ from $b(w)$ to $s(w)$ and $P(s(w),b(w))$ from $s(w)$ to $b(w)$ such that $P(s(w),b(w))$ avoids $P(b(w),s(w))$. Finally, $P(w,s(w))$ is congruent to $P(u,b(w))$ and the former avoids the latter. The circular $N$ is formed by the walks
\begin{align*}
    X &= P(w,s(w)) P(s(w),b(w)) P(u,b(w))^{-1} P(u,s(u)) P(s(u),b(u)) P(w,b(u))^{-1},\\
    Y &= P(u,b(w)) P(b(w),s(w)) P(w,s(w))^{-1} P(w,b(u)) P(b(u),s(u)) P(u,s(u))^{-1},\\
\end{align*}
and $Z$ is the sub-walk of $Y$ from $u$ (starts at the bottom of the figure) to $a$, then the edge $ab$, and the sub-walk of $X$ from $b$ to $w$ ($w$ at the top of the figure).
\qed

In \cite{eklt} the authors prove an L versus NL dichotomy for undirected graphs, giving a description of the induced subgraphs responsible for the list homomorphism problem being NL-hard. It is not difficult to find a circular $N$ in each of these subgraphs. For instance, in a reflexive four-cycle with consecutive vertices $a,b,c,d$, we can take the walks $a,b,c,d,a$, and $b,c,d,a,b$, and $b,c,c,d,a$. Therefore Theorem~\ref{main_thm} generalizes the L versus NL dichotomy from \cite{eklt}.

We turn our attention to those digraphs for which the list homomorphism problem is definable in first-order logic.
\begin{dfn}[\cite{adrien}]
	Let $H$ be a digraph. We say that the edges $ab,cd \in E(H)$ (both forward or both backward) are \emph{independent}, if neither $ad$ nor $cb$ is a (forward or backward, respectively) edge of $H$. A \emph{bicycle} in $H$ consists of two walks $X = x_0,x_1,\dots,x_n$ and $Y = y_0,y_1,\dots,y_n$ where all edges are forward, $X$ avoids $Y$, and $y_ix_{i+1}$ is a forward edge for each $i = 0,1,\dots,n$ (where the indices of the variables are modulo $n$).
\end{dfn}

\begin{tm}[\cite{adrien}]
    Let $H$ be a digraph. Then LHOM$(H)$ is definable in first-order logic if and only if $H$ contains neither a pair of independent edges nor a bicycle.
\end{tm}

Again, as expected, a circular $N$ either contains a pair of independent edges or a bicycle.
\begin{prp}\label{cN->bicycle+not_tele}
  Let $H$ be a digraph that contains a circular $N$. Then $H$ contains either a pair of independent edges or a bicycle.
\end{prp}
\pf\footnote{The proof of the proposition uses an argument similar to the proof of Proposition~2.20 in \cite{adrien}.}
	Assume that the circular $N$ consists of the walks $X = x_0,x_1,\dots,x_n$, $Y = y_0,y_1,\dots,y_n$ and $Z$ (we do not need $Z$). Recall that $X$ avoids $Y$. Assume that for some $i$, there is no faithful edge $y_ix_{i+1}$. Then the edges $x_ix_{i+1}, y_iy_{i+1}$ are independent. If $X$ and $Y$ do not give a bicycle, then there is some $j$ such that $x_jx_{j+1}$, $y_jy_{j+1}$, $y_jx_{j+1}$, $x_{j+2}x_{j+1}$, $y_{j+2}y_{j+1}$, $x_{j+2}$, $y_{j+1}$ are all forward (or backward) edges. The edges $x_jx_{j+1}$ and $y_{j+2}y_{j+1}$ are independent.
\qed

We now summarize the known complexity classification results about LHOM$(H)$, where $H$ is a digraph. The following theorem also uses results from \cite{soda,lalota,lartes,adrien}.

\begin{tm}\label{summary}
Let $H$ be a digraph.
\begin{enumerate}
  \item If $H$ contains a DAT, then LHOM($H$) is NP-complete.
  \item If $H$ contains no DAT but $H$ contains a circular $N$, then LHOM($H$) is in P but is hard for NL.\label{case_2}
  \item If $H$ contains no circular $N$ but contains a bicycle or a pair of independent edges, then LHOM($H$) is in L but is hard for L (under first-order reductions).
  \item If $H$ contains no bicycle and no pair of independent edges, then LHOM($H$) is definable in first-order logic.
\end{enumerate}
\end{tm}

\emph{Remark:} Using Proposition~\ref{cN->bicycle+not_tele} and the simple observation that a bicycle contains a circular $N$, we can conclude that if $H$ contains no pair of independent edges, then the following are equivalent: $(i)$ $H$ has no circular $N$; $(ii)$ $H$ has no bicycle; $(iii)$ LHOM$(H)$ is definable in first-order logic.

Note that a polynomial time algorithm is known to find a DAT if one exists \cite{soda}. We conclude the paper by giving a polynomial time algorithm to find a circular $N$ if one exists.

\begin{tm} \label{circN-recog}
There is an algorithm that decides in time polynomial in $|V(H)|$ whether $H$ contains a circular $N$.
\end{tm}

\pf
Given a digraph $H$, we define a {\em coloured triple digraph} $H^{++}$ with vertex set
$V(H) \times V(H) \times V(H)$, in which there is a forward edge from $(a,b,c)$ to $(a',b',c')$
exactly when $H$ has edges $aa', bb', cc'$, does not have the edge $ac'$, and does not have
the edge $ab'$ or does not have the edge $bc'$, with all these edges or non-edges of $H$
being in the same direction, forward or backward. The colour of the edge $(a,b,c)(a',b',c')$
is green if both edges $ab', bc'$ are missing, {\em blue} if only edge the $ab'$ is missing, and
{\em brown} if only the edge $bc'$ is missing. It is now easy to observe that $H$ has a circular
$N$ if and only if $H^{++}$ contains a forward directed closed walk from $(x,y,y)$ to $(x,x,y)$ for some $x$ and $y$
in which no brown edge precedes a blue edge. For a fixed $x, y$, this can be checked by finding
all vertices $(a,b,c)$ reachable from $(x,y,y)$ on directed paths without brown edges, and all
vertices $(a,b,c)$ that can reach $(x,x,y)$ on directed paths without blue edges; if these two sets
intersect, there is a forward directed closed walk from $(x,y,y)$ to $(x,x,y)$ in which no brown edge
precedes a blue edge, and hence $H$ contains a circular $N$. This check takes time polynomial in the size of $H$.
\qed

\subsection{Open Problems}

It would be an important step towards understanding the complexity of CSPs to verify the following statement: LHOM($H$) is either in NL or is hard for P. That is, we would like to refine the complexity characterisation of those list homomorphism problems that belong to case~\ref{case_2} of Theorem~\ref{summary}. We note that both digraphs for which LHOM($H$) is NL-complete, and digraphs for which LHOM($H$) is P-complete are known to exist.

We also ask if it is possible to generalize our results to other relational structures, i.e., to show that a conservative CSP is in logspace, or is hard for a complexity class containing L, e.g., NL or Mod$_p$L (where $p$ is a prime).

Finally, if $H$ is a symmetric digraph (i.e., an undirected graph) that contains no circular $N$, then it is known that LHOM($H$) is definable in symmetric Datalog (note that symmetric Datalog programs can be evaluated in logspace, \cite{elt}). Can we show that if $H$ is digraph that contains no circular $N$, then LHOM($H$) is in symmetric Datalog? 
\bibliographystyle{abbrv}
\bibliography{bibliography}

\end{document}